\documentclass[reprint,superscriptaddress,amssymb,amsmath,aps,showpacs,10pt,floatfix,prb,longbibliography]{revtex4-2}
\def\CVS{CsV$_3$Sb$_5$}
\def\KVS{KV$_3$Sb$_5$}
\def\RVS{RbV$_3$Sb$_5$}
\def\AVS{$A$V$_3$Sb$_5$}
\def\cm{cm$^{-1}$}
\def\Tc{$T_{\rm CDW}$}

\usepackage{graphicx}%
\usepackage{color}
\usepackage{epstopdf}
\usepackage{amssymb}
\usepackage{amsmath}
\usepackage{amsfonts}
\usepackage[note-name=, use-sort-key = false]{notes2bib}
\usepackage{xcolor}

\usepackage{color}
\usepackage[colorlinks,bookmarks=false,citecolor=darkblue,linkcolor=red,urlcolor=blue]{hyperref} 
\definecolor{darkred}{rgb}{0.7,0.0,0.0}

\definecolor{darkblue}{rgb}{0,0.02,0.45}

\definecolor{darkgreen}{rgb}{0.02,0.45,0.0}

\definecolor{violet}{rgb}{0.8,0.2,0.6}

\begin{document}
\title{Optical study of \RVS: Multiple density-wave gaps and phonon anomalies}

\author{M. Wenzel}
\affiliation{1. Physikalisches Institut, Universit{\"a}t Stuttgart, 70569
Stuttgart, Germany}

\author{B. R. Ortiz}
\affiliation{Materials Department and California Nanosystems Institute,
University of California Santa Barbara, Santa Barbara, CA, 93106, United States}
\affiliation{Materials Department, University of California Santa Barbara, Santa Barbara, CA, 93106, United States}

\author{S. D. Wilson}
\affiliation{Materials Department, University of California Santa Barbara, Santa Barbara, CA, 93106, United States}

\author{M. Dressel}
\affiliation{1. Physikalisches Institut, Universit{\"a}t Stuttgart, 70569
Stuttgart, Germany}

\author{A. A. Tsirlin}
\email{altsirlin@gmail.com}
\affiliation{Experimental Physics VI, Center for Electronic Correlations and Magnetism, University of Augsburg, 86159 Augsburg, Germany}

\author{E. Uykur}
\email{ece.uykur@pi1.physik.uni-stuttgart.de}
\affiliation{1. Physikalisches Institut, Universit{\"a}t Stuttgart, 70569
Stuttgart, Germany}
\affiliation{Helmholtz Zentrum Dresden Rossendorf, Inst Ion Beam Phys \& Mat Res, D-01328 Dresden, Germany}
\date{\today}
%\keywords{}

\begin{abstract}
Temperature-dependent reflectivity studies on the non-magnetic kagome metal \RVS\ in a broad energy range (50~\cm\ -- 20000~\cm, equivalent to 6~meV -- 2.5~eV) down to 10~K are reported. Below $T_{\rm CDW}=102$~K, the optical spectra demonstrate a prominent spectral-weight transfer from low to higher energies as the fingerprint of the charge-density wave (CDW) formation with the opening of a partial gap. A detailed analysis reveals two energy scales of, respectively, $\sim$~800~\cm\ (100~meV) and 360~\cm\ (45~meV), the latter visible below 50~K only. Additionally, two modes at, respectively, 160~\cm\ (20~meV) and 430~\cm\ (53~meV) can be traced both above and below \Tc. They show strong anomalies already above \Tc\ with a further renormalization across the transition, suggesting the importance of the electron-phonon coupling in \RVS\ in both normal and CDW states. While the 160~\cm\ mode can be attributed to the E$_{1u}$ phonon, the 430~\cm\ mode could not be reproduced in our phonon calculations. The antiresonance nature of this mode suggests a nontrivial electron-phonon coupling in \RVS. A distinct localization peak observed at all temperatures signals damped electron dynamics, whereas the reduced Drude spectral weight manifests moderate deviations from the band picture in \RVS.
\end{abstract}

\maketitle

\section{Introduction}
Kagome metals became model compounds for studying effects of correlations along with topologically non-trivial electronic states \cite{Liu2019}. Driven by the spatially separated metallic kagome planes, electronic structures of kagome metals feature flat bands and linearly dispersing topological Dirac bands, as has been shown in several magnetic systems of recent interest \cite{Ye2018, Lin2018, Yin2019, Kang2020}.

The non-magnetic \AVS\ (A = K, Rb, Cs) series opens up new opportunities to study electronic properties of the kagome metals \cite{Ortiz2019}. These compounds crystallize in the $P6/mmm$ space group with V atoms forming a kagome lattice and Sb1 atoms filling the centers of the hexagons. The kagome networks are stacked along the $c$-axis and separated by Sb2 honeycomb layers and A alkali atoms. This results in an overall quasi-2D structure with well-isolated kagome planes. All key features of a nearest-neighbor kagome metal -- linear bands, saddle points, and flat bands -- can be indeed distinguished in the electronic structures of \AVS~\cite{Neupert2021}.

Transport and magnetic measurements on \AVS\ revealed a strong anomaly at \Tc\ = 102 K for \RVS\ \cite{Yin2021} (78 K for \KVS\ \cite{Ortiz2019} and 94 K for \CVS\ \cite{Ortiz2020}). Moreover, \RVS\ becomes superconducting at $T_{\mathrm{c}}=0.92\,\mathrm{K}$ \cite{Yin2021} (0.93 K for \KVS\ \cite{Ortiz2021} and 2.5 K for \CVS\ \cite{Ortiz2020}). While the exact nature of the high-temperature anomaly is still under debate, the robust nature of this transition under external magnetic fields suggests charge-density wave (CDW) as the plausible origin. Indeed, band saddle points (van Hove singularities) observed in \AVS\ in the vicinity of the Fermi level~\cite{Nakayama2021,Liu2021,Kang2021} should lead to a CDW instability of a kagome metal~\cite{Kiesel2013,Park2021,Denner2021}. The resulting CDW state is quite unusual, as it features large anomalous Hall effect~\cite{Yang2020, Yu2021}, multiple energy gaps~\cite{Nakayama2021, Luo2021, Wang2021d}, and intrinsic chirality~\cite{Yu2021, Mielke2021, Shumiya2021}. One general and hitherto unresolved issue is the clear distinction between the bulk and surface effects in the CDW state of \AVS. For example, two different coexisting superstructures were observed in surface-sensitive measurements of \CVS~\cite{Zhao2021b}, but only one of them is believed to occur in the bulk~\cite{Xiang2021,Ptok2021}. A temperature-driven re-arrangement of the CDW order has been proposed in \CVS\ as well~\cite{Stahl2021}.

%%%%%%%%%%%%%%%%%%%%%%%%%%%%%%%%%%%%%%%%%%%%%%%%%%%%%%%%%%%%%
\begin{figure}
	\centering
	\includegraphics[width=1\columnwidth]{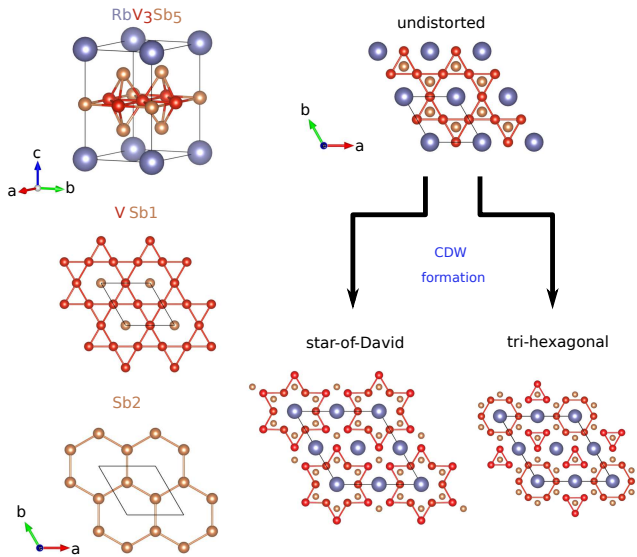}
	\caption{Crystal structure of \RVS. V-network is stabilized with the Sb1 atoms. The V-Sb1 kagome net is separated by the Sb2 honeycomb sheets and Rb atoms. Two possible structural distortions below \Tc\ are the star-of-David and tri-hexagonal structures~\cite{Uykur2021a, Uykur2022}.  }		
	\label{structure}
\end{figure}
%%%%%%%%%%%%%%%%%%%%%%%%%%%%%%%%%%%%%%%%%%%%%%%%%%%%%%%%%%%%

Despite many similarities shared by all three compounds -- \KVS, \RVS, and \CVS\ -- they feature somewhat different energy scales and, possibly, different underlying interactions. Density-functional-theory (DFT) band-structure calculations could perfectly reproduce the optical response of \CVS~\cite{Uykur2021a}, but not of \KVS~\cite{Uykur2022}, suggesting that deviations from the band picture may be enhanced across the series as the size of the alkaline metal is reduced. This motivates a comparative optical study of \RVS\ as the possible intermediate case.

Below, we show that this compound is indeed intermediate in terms of its bands near the Fermi level and their saddle points. We further confirm that several pertinent features -- hindered electron dynamics witnessed by the low-energy localization peak, and strong electron-phonon coupling revealed by the broadening of the phonon modes -- are common across the \AVS\ series. Additionally, we are able to detect two distinct CDW gaps in \RVS. These gaps are integral to the putative topological state but could be previously observed with surface-sensitive techniques only~\cite{Cho2021, Liu2021}. Our optical study confirms their bulk nature, thus setting a benchmark for the CDW state of \AVS.

%%%%%%%%%%%%%%%%%%%%%%%%%%%%%%%%%%%%%%%%%%%%%%%%%%%%%%%%%%%%%
\begin{figure*}
	\centering
	\includegraphics[width=2\columnwidth]{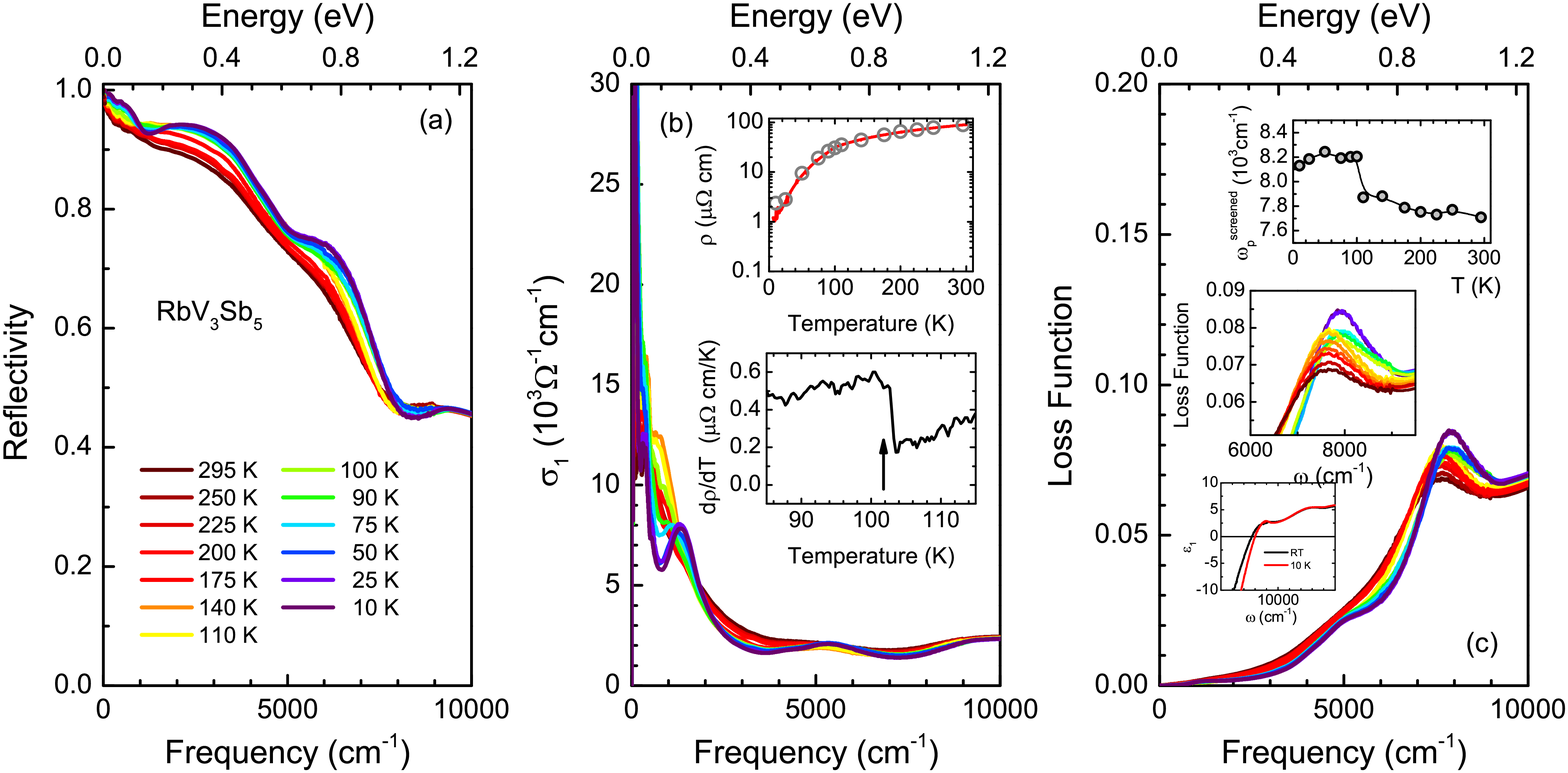}
	\caption{(a) Temperature-dependent reflectivity over a broad frequency range measured in the $ab$-plane. (b) Calculated real part of the optical conductivity. The inset shows the resistivity values obtained from the Hagen-Rubens fit of the reflectivity overlapped with the dc resistivity measurement. At 102 K, a kink in the resistivity marks the CDW transition that is better visible in the first derivative. (c) Temperature dependence of the dielectric loss function. The position of the maximum indicates the screened plasma frequency, with temperature dependence given in the inset. A clear increase is observed across the CDW transition at \Tc = 102~K. This change can also be seen in the frequency-dependent dielectric permittivity at room temperature and at 10~K shown in another inset of the same panel. The high-energy limit is taken as $\varepsilon_\infty$.}		
	\label{Ref_OC}
\end{figure*}
%%%%%%%%%%%%%%%%%%%%%%%%%%%%%%%%%%%%%%%%%%%%%%%%%%%%%%%%%%%%

\section{Methods}
\subsection{Experimental}
High-quality single crystals were prepared as explained in Refs.~\cite{Ortiz2019, Ortiz2021}. For the optical measurements, a sample with the dimensions of $\sim 1.5 \times 1.5 $~mm$^2$ surface area was used, while the thickness of the specimen was around 0.2~mm. The sample was freshly cleaved prior to the optical experiments. On the same crystal, four-point resistivity measurements were performed in order to determine the CDW transition temperature and confirm the stoichiometry. The kink at 102~K marking the CDW transition [Fig.~\ref{Ref_OC}(b), inset], as well as the overall behaviour agrees well with the previous reports \cite{Yin2021}.

Temperature-dependent reflectivity measurements in the $ab$-plane over a broad frequency range from 50 to  20000 \cm\ (6 meV -- 2.5 eV) were performed down to 10~K. While for the high-energy range ($\omega\,>\,600\,\mathrm{cm^{-1}}$) a Bruker Vertex 80v spectrometer with an incorporated Hyperion IR microscope was used, the low-energy range was measured with a Bruker IFS113v spectrometer and a custom-built cryostat. Freshly evaporated gold mirrors served as reference in these measurements. The absolute value of the reflectivity was obtained by an in-situ gold-overcoating technique in the far-infrared range, as described in Ref.~\cite{Homes1993}.

Below 50 \cm, we use standard Hagen-Rubens extrapolation, considering the metallic nature of our sample, while for the high-energy range we utilize x-ray scattering functions for extrapolating the data \cite{Tanner2015}. The optical conductivity is then calculated from the measured reflectivity via Kramers-Kronig analysis.

\subsection{Computational}
Density-functional-theory (DFT) calculations of the band structure and optical conductivity were performed in the \texttt{Wien2K}~\cite{wien2k,Blaha2020} code using Perdew-Burke-Ernzerhof flavor of the exchange-correlation potential~\cite{pbe96}. We used experimental structural parameters from Ref.~\cite{Ortiz2019} for the undistorted \RVS{} structure, whereas possible CDW structures were obtained by a structural relaxation in \texttt{VASP}~\cite{vasp1,vasp2} similar to Ref.~\cite{Uykur2021a}. Spin-orbit coupling was included for the calculations of band structure and optical conductivity.

Self-consistent calculations and structural relaxations were converged on the $24\times 24\times 12$ $k$-mesh for the undistorted \RVS\ structure (normal state) and $12\times 12\times 12$ $k$-mesh for the distorted structures (CDW). Optical conductivity was calculated on the $k$-mesh with up to $100\times 100\times 50$ points for the normal state and $36\times 36\times 36$ points for the CDW states.

\section{Results and Discussion}

\subsection{Optical spectra}
\begin{figure}
	\centering
	\includegraphics[width=1\columnwidth]{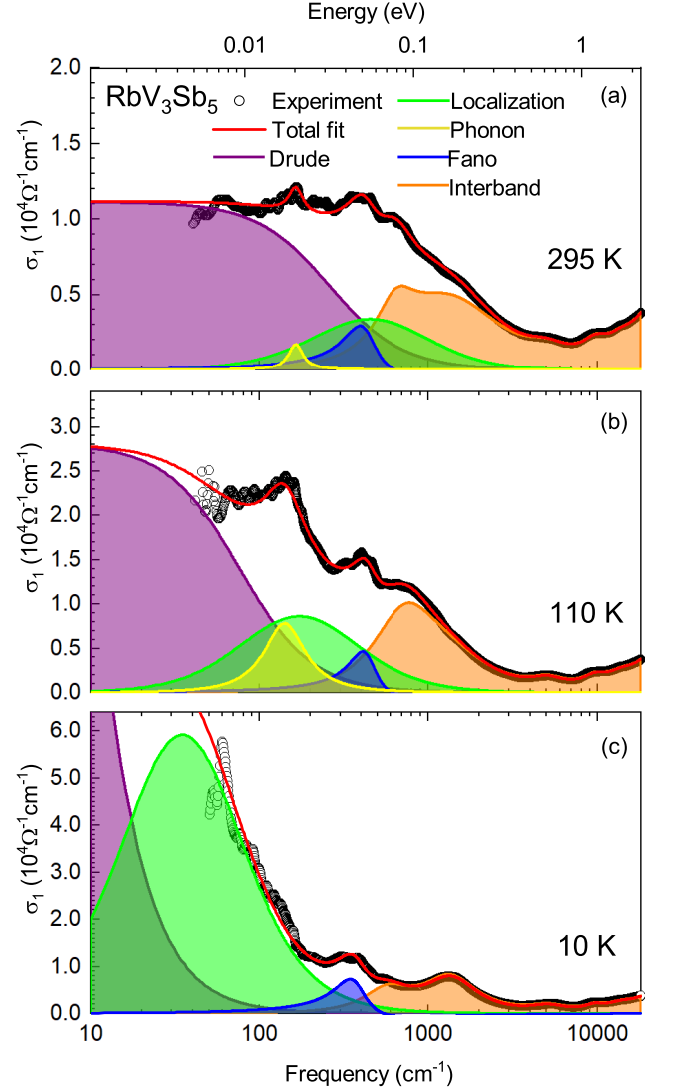}
	\caption{Decomposition of the optical conductivity at (a) room temperature, (b) 110~K $>$ \Tc, and (c) 10~K $<$ \Tc\ consisting of a Drude peak (purple), a localization peak (green), a phonon (yellow), a Fano resonance (blue), and multiple interband transitions (orange). Note that at 10~K only the tail of the localization peak is visible, but it is sufficient to estimate the peak position.}			\label{fit300K}
\end{figure}

Fig.~\ref{Ref_OC} (a) displays the temperature-dependent reflectivity of \RVS. The high reflectivity values at low frequencies, as well as the Drude-like increase in the optical conductivity [Fig.~\ref{Ref_OC} (b)] demonstrate the metallic nature of the sample. Conductivity values in the $\omega\,\rightarrow\,0$ limit are obtained from the Hagen-Rubens fit of the reflectivity {[Fig.~\ref{fitting}(a)] and match well with the four-probe dc resistivity measurements performed on the same sample, as shown in the inset of Fig.~\ref{Ref_OC} (b). Upon going through the CDW transition, significant changes in the optical properties occur. Around 0.15~eV, a dip develops in the reflectivity, echoed by a spectral weight transfer to an additional peak at around 0.17~eV in the optical conductivity below \Tc. The low-energy optical conductivity is highlighted in Fig.~\ref{fitting} (b), where the spectral-weight transfer to higher frequencies is shown by the green arrow. The solid circles in the figure represent the dc conductivity values, whereas the orange arrow highlights the Fano-resonance. This Fano resonance is one of the signatures of the strong electron-phonon coupling that will be further discussed in Section~\ref{sec:Fano}. 

The CDW transition also affects the free-carrier dynamics, as seen from the evolution of the screened plasma frequency, $\omega_{\mathrm{p}}^{\mathrm{screened}}$, which can best be estimated from the position of the maximum in the dielectric loss function, $-{\rm Im}\{1/\tilde{\varepsilon}\}$ \cite{Dressel2002}. The real plasma frequency is masked by the interband transitions, but it can be calculated as $\omega_{\mathrm{p}}\,=\,\omega_{\mathrm{p}}^{\mathrm{screened}}\,\cdot\,\sqrt{\varepsilon_{\infty}}$ where we use the temperature-independent $\varepsilon_{\infty}=6$ estimated from the high-energy limit of the permittivity, as shown in the inset of Fig.~\ref{Ref_OC}(c). Given this simple relation between $\omega_p$ and $\omega_{\mathrm{p}}^{\mathrm{screened}}$, the abrupt increase around \Tc\ is expected in the real plasma frequency, too.

As the plasma frequency can also be given as $\omega^2_{\mathrm{p}}\propto n/m^*$ ($n$ is the carrier density and $m^*$ is the effective mass), the overall increase in $\omega_p$ can be attributed to either increasing carrier density or decreasing effective mass. Both scenarios are possible in \RVS. The CDW transition in \RVS\ gaps out some of the electronic bands~\cite{Liu2021,Cho2021}, thus affecting $m^*$. Concurrently, Hall effect measurements on \KVS~\cite{Yang2020} and \CVS~\cite{Yu2021} suggest that below \Tc\ the carrier density increases.

\subsection{Decomposition}
Different contributions to the optical spectra are modeled with the Drude-Lorentz approach,
\begin{equation}
\label{Eps}
\tilde{\varepsilon}(\omega)= \varepsilon_\infty - \frac{\omega^2_{p,{\rm Drude}}}{\omega^2 + i\omega/\tau_{\rm\, Drude}} + \sum\limits_j\frac{\Omega_j^2}{\omega_{0,j}^2 - \omega^2-i\omega\gamma_j} .
\end{equation}
Here, $\omega_{p,{\rm Drude}}$ and $1/\tau_{\rm\,Drude}$ are the plasma frequency and the scattering rate of the itinerant carriers, respectively. The parameters $\omega_{0,j}$, $\Omega_j$, and $\gamma_j$ describe the resonance frequency, width, and the strength of the $j^{th}$ excitation, respectively. Finally, $\varepsilon_\infty$ stands for the high-energy contributions to the real part of the dielectric permittivity [$\tilde{\varepsilon}=\varepsilon_1 + i\varepsilon_2$].

In addition to the classical Lorentzian and Drude contributions, a Fano-like shaped peak is observed around 400 \cm\ along with the strongly temperature-dependent absorption feature that appears around 500~\cm\ at room temperature and systematically shifts to lower energies upon cooling. In line with the previous optical studies of the kagome metals, this latter feature is assigned to a so-called localization peak \cite{Uykur2022, Uykur2021a, Biswas2020} that signals hindered electron dynamics. The nature and temperature evolution of this peak are further discussed in Sec.~\ref{sec:localization} below.

%%%%%%%%%%%%%%%%%%%%%%%%%%%%%%%%%%%%%%%%%%%%%%%%%%%%%%%%%%%%%
\begin{figure}
	\centering
	\includegraphics[width=1\columnwidth]{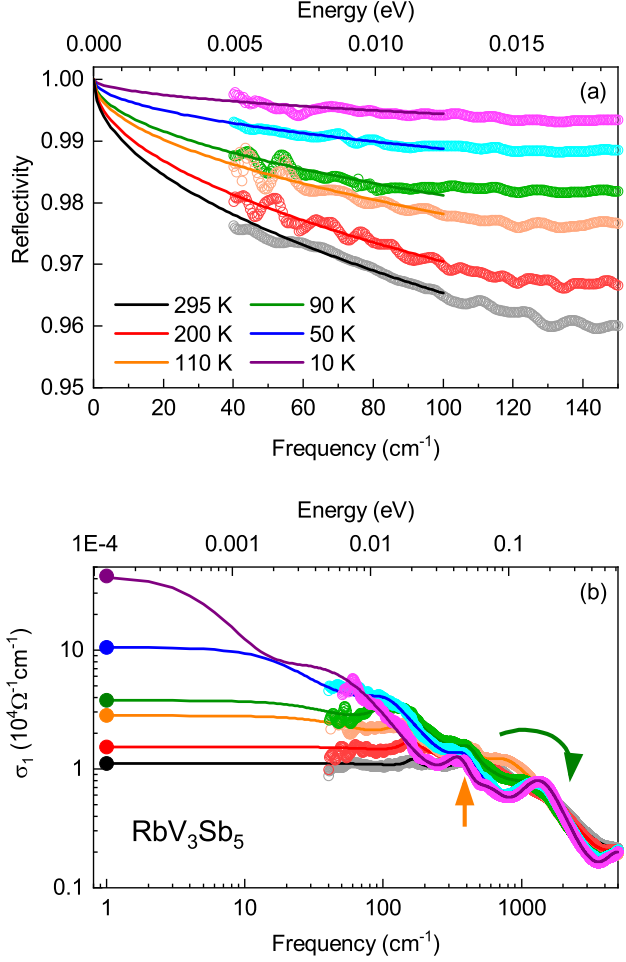}
	\caption{(a) Low-energy reflectivity and the corresponding Hagen-Rubens fits at selected temperatures. (b) Low-energy optical conductivity shown for the same selected temperatures. The solid lines are the fits to the experimental spectra as described in the text. While at 10~K the localization peak lies outside of the measurement range, its position can be estimated assuming linear temperature dependence and using the high-energy tail, which is still visible in our data. The solid circles are the fixed dc conductivity values obtained from the Hagen-Rubens fits. The orange arrow marks the Fano resonance, and the green arrow illustrates the spectral weight transfer to an additional peak that appears below \Tc.}		
	\label{fitting}
\end{figure}
%%%%%%%%%%%%%%%%%%%%%%%%%%%%%%%%%%%%%%%%%%%%%%%%%%%%%%%%%%%%

Total dielectric permittivity takes the form
\begin{equation}
\tilde{\varepsilon}(\omega) = \tilde{\varepsilon}_{\rm Drude}(\omega) + \tilde{\varepsilon}_{\rm Lorentz}(\omega)+\tilde{\varepsilon}_{\rm local}(\omega) + \tilde{\varepsilon}_{\rm Fano}(\omega) .
\end{equation}
The complex optical conductivity [$\tilde{\sigma}=\sigma_1 + i\sigma_2$] is then calculated as
\begin{equation}
\tilde{\sigma}(\omega)= -i\omega[\tilde{\varepsilon} (\omega) -\varepsilon_\infty]/4\pi \quad .
\label{Cond}
\end{equation}

Examples of the decomposed spectra are given in Fig.~\ref{fit300K}. The narrowing of the Drude contribution along with the red shift of the localization peak are clearly visible in the spectra. Furthermore, the overall sharpening of the interband transitions down to 110~K and the redistribution of the spectral weight at \Tc\ are demonstrated.}

\subsection{Localization Peak}
\label{sec:localization}

%%%%%%%%%%%%%%%%%%%%%%%%%%%%%%%%%%%%%%%%%%%%%%%%%%%%%%%%%%%%%
%%%%%%%%%%%%%%%%%%%%%%%%%%%%%%%%%%%%%%%%%%%%%%%%%%%%%%%%%%%%

\begin{figure}
	\centering
	\includegraphics[width=1\columnwidth]{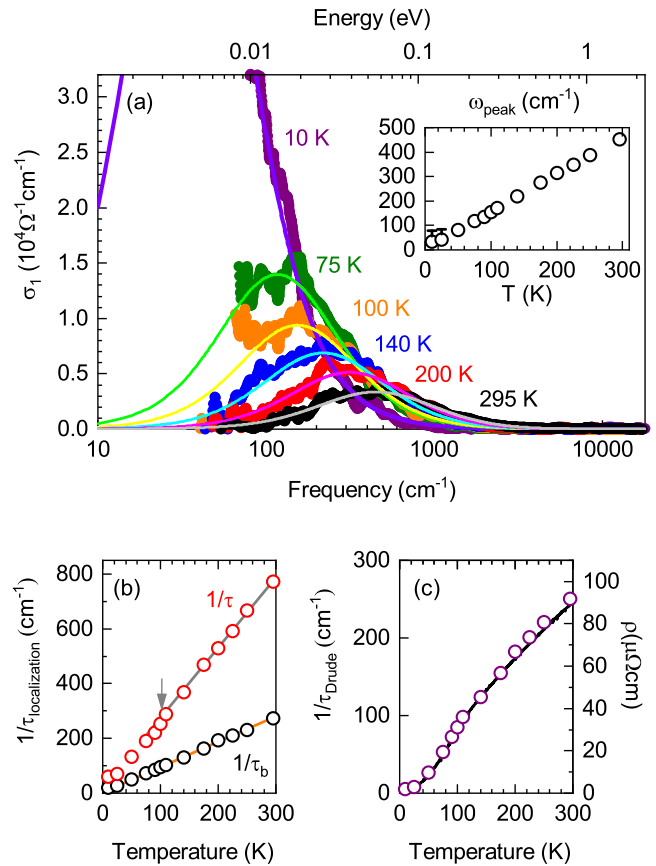}
	\caption{(a) Temperature dependence of the localization peak obtained after subtracting all other contributions from the experimental spectra. The inset shows the temperature evolution of the peak position. The error bars at 10~K and 25~K arise from the fact that the peak center lies outside of the frequency range of our measurement. (b) Elastic scattering ($\tau$) and back scattering ($\tau_{\mathrm{b}}$) of the localization peak of \RVS. (c) Scattering rate of the Drude component overlayed with the dc resistivity.}  		
	\label{scattering}
\end{figure}

Having described the general decomposition of the optical spectrum, we now turn to a more detailed analysis of the localization peak. Its temperature evolution is plotted in Fig.~\ref{scattering} (a), after subtracting the Drude, phonon, and interband contributions from the experimental data. The strong and linear red-shift of the peak position upon cooling, given in the inset of Fig.~\ref{scattering} (a), makes this feature clearly distinguishable from the interband transitions. Furthermore, the interband transitions are well reproduced by DFT calculations, as discussed in Sec.~\ref{calcsection}. The absence of this peak in the calculated spectrum gives further evidence for its interpretation in terms of an intraband process.

Localization, or displaced Drude peaks in the optical conductivity can have different nature and have been reported for transition-metal oxides \cite{Kostic1998, *Lee2002, *Bernhard2004, *NanlinWang2004, *Rozenberg1995, *Jonsson2007, *Takenaka2002, *Jaramillo2014}, cuprate superconductors \cite{Puchkov1995, *Hwang2007, *Tsvetkov1997,  *Osafune1999, *Uykur2011}, and organic conductors \cite{Dong1999, *Takenaka2005, *Fratini2021}. It has been shown that the partial localization of the charge carriers shifts the Drude peak to finite frequencies. Different mechanisms of this localization can be envisaged, including electron-phonon interactions, electronic correlations, as well as localization of carriers caused by a structural disorder in the material.

Among the different theoretical frameworks that have been developed for the localization peak~\cite{Smith, Hartnoll}, we choose the model of a displaced Drude peak by Fratini \textit{et al.} \cite{Fratini2014}, where the classical Drude response is modified with the backscattering of the electrons, leading to a shift of the zero-frequency response to a finite value:
\begin{equation}
\begin{split}
\tilde{\sigma}_{\rm local}(\omega)= & \frac{C}{\tau_{\mathrm{b}}-\tau}\frac{\tanh\{\frac{\hbar\omega}{2k_{\mathrm{B}}T}\}}{\hbar\omega}\,\cdot\, \\
&\mathrm{Re}\left\{\frac{1}{1-\mathrm{i}\omega\tau}-\frac{1}{1-\mathrm{i}\omega\tau_{\mathrm{b}}}\right\} \quad .
\end{split}
\end{equation}

Here, $C$ is a constant, $\hbar$ is the reduced Planck constant, $k_{\mathrm{B}}$ the Boltzmann constant. Furthermore, $\tau$ stands for the elastic scattering time of the standard Drude model, whereas $\tau_b$ is the backscattering of the electrons due to localization effects. Here one should also point out that $\tau_b > \tau$ with a longer timescale. The formalism, as presented in Ref.~\onlinecite{Fratini2014}, assumes possible localization effects, due to interactions of charge carriers with low-energy degrees of freedom, such as phonons, electric or magnetic fluctuations, which lead to a backscattering of the electrons.

For a further insight, we plotted the elastic scattering and the backscattering of the localization peak in Fig.~\ref{scattering} (b). As noted by the arrow, the elastic scattering shows a slight change across the CDW transition. This change corresponds to a partial gapping of the Fermi surface~\cite{Liu2021,Cho2021} and the change in the carrier density across \Tc. On the other hand, the backscattering mechanism seems to be unaffected by the CDW formation. In Fig.~\ref{scattering} (c), we have also shown the elastic scattering of the Drude component, which is overlayed with the dc resistivity. The remarkably similar temperature evolution suggests that the dc transport is mainly governed by the Drude component and not by the incoherent localization peak.

%%%%%%%%%%%%%%%%%%%%%%%%%%%%%%%%%%%%%%%%%%%%%%%%%%%%%%%%%%%%%
%%%%%%%%%%%%%%%%%%%%%%%%%%%%%%%%%%%%%%%%%%%%%%%%%%%%%%%%%%%%

Regardless of the exact microscopic interpretation of this localization peak, a striking feature of its temperature evolution is the insensitivity to \Tc. The effects that hinder electron dynamics seem to be purely thermal in nature and vanish in the $T\rightarrow 0$ limit. A similar behavior is observed in \CVS~\cite{Uykur2021a}, but not in \KVS\ where the localization peak also shifts toward low energies upon cooling, yet it saturates at around 300~\cm\ and does not reach zero energy even at 10~K~\cite{Uykur2022}.

\subsection{Energy scale of the charge-density wave}

One of the key features of the \AVS\ family is the formation of a density-wave state accompanied by a superstructure \cite{Ortiz2020}. While there is a consensus on the $2\times2$ in-plane modulation \cite{Jiang2021, Liang2021, Li2021a, Chen2021b, Shumiya2021, Wang2021b}, both two-fold and four-fold modulations along the $c$ axis have been proposed \cite{Chen2021b, Zhao2021b, Ortiz2021a, Luo2021b}. This superstructure formation is concomitant with the reduction in the density of states at the Fermi level. Recent quantum oscillation \cite{Ortiz2021a} and ARPES \cite{Nakayama2021, Liu2021} studies suggested that vanadium bands around the $M$ point become gapped, in line with theoretical proposals of band saddle points at $M$ as the main origin of the density-wave instability \cite{Feng2021, Park2021}. The presence of multiple $d$-bands in \AVS~\cite{Neupert2021} implies that this instability can have more than one energy scale revealed by several gap features that were indeed detected spectroscopically, but only with surface-sensitive techniques, so far \cite{Nakayama2021, Liu2021, Wang2021d}.

\begin{figure}
	\centering
	\includegraphics[width=1\columnwidth]{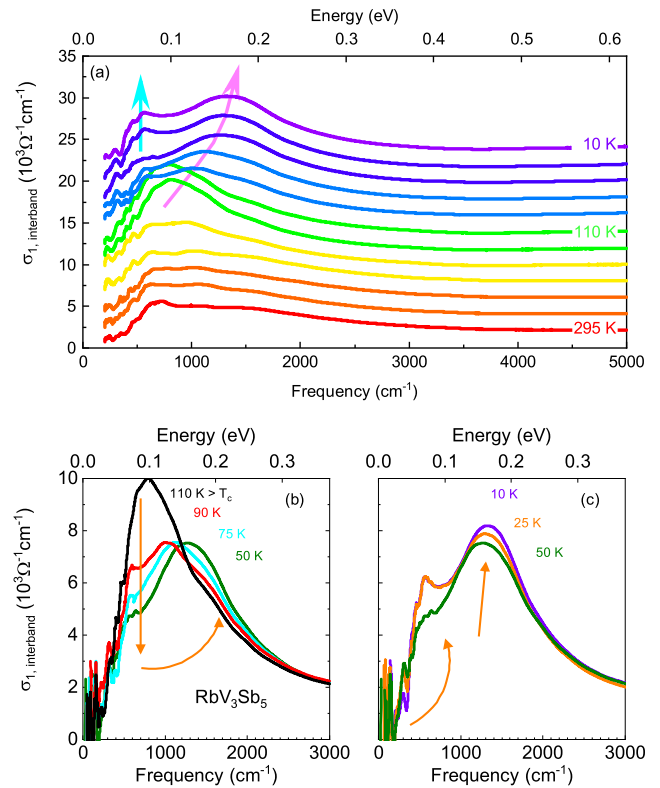}
	\caption{(a) Temperature-dependent interband transitions. The curves are shifted by 2000 $\Omega^{-1}$cm$^{-1}$ for clarity. The gradual sharpening of the interband transitions on cooling is interrupted with the CDW formation that manifests itself by the shift of the spectral weight toward higher energies (pink arrow). Below 50~K, an additional absorption appears around 50~meV and can be most clearly seen in the non-shifted spectra of panel (c). (b) Interband transitions of \RVS\ as a function of temperature below \Tc\ down to 50~K. The 110~K data are shown as reference for the change across \Tc. (c) Interband transitions as a function of temperature below 50~K show the emergence of the second energy scale at low temperatures. The arrows in (a) and (b) demonstrate the SW transfer across \Tc\ and below 50~K.  }		
	\label{interband}
\end{figure}

Here, we show that these multiple gaps can be corroborated by the bulk optical probe. Fig.~\ref{interband}(a) displays temperature dependence of the interband transitions and demonstrates the sharpening of the absorption on cooling down to \Tc. Below \Tc, spectral weight shifts to higher energies (pink arrow), indicating the opening of a CDW gap. 
%with decreasing temperature and and an abrupt change in behavior below 110~K as demonstrated with the arrows. A conventional spectral weight transfer from low to higher energies, as expected in the CDW formation. This spectral weight transfer from the peak at 70~meV is observed down to 10~K [Fig.~\ref{interband}(a) and (b)]. 
A very different behavior is seen at even lower temperatures, below 50~K, where the low-energy spectral weight around 50\,meV starts growing again [Fig.~\ref{interband}(b)] because of the redistribution from even lower energies (cyan arrow). A convenient way to trace these spectral weight transfers is by using the spectral weight ratio below and above \Tc, similar to, e.g., iron pnictides \cite{Schafgans2012}.

%%%%%%%%%%%%%%%%%%%%%%%%%%%%%%%%%%%%%%%%%%%%%%%%%%%%%%%%%%%%%
%%%%%%%%%%%%%%%%%%%%%%%%%%%%%%%%%%%%%%%%%%%%%%%%%%%%%%%%%%%%
%%%%%%%%%%%%%%%%%%%%%%%%%%%%%%%%%%%%%%%%%%%%%%%%%%%%%%%%%%%%%
\begin{figure}
	\centering
	\includegraphics[width=1\columnwidth]{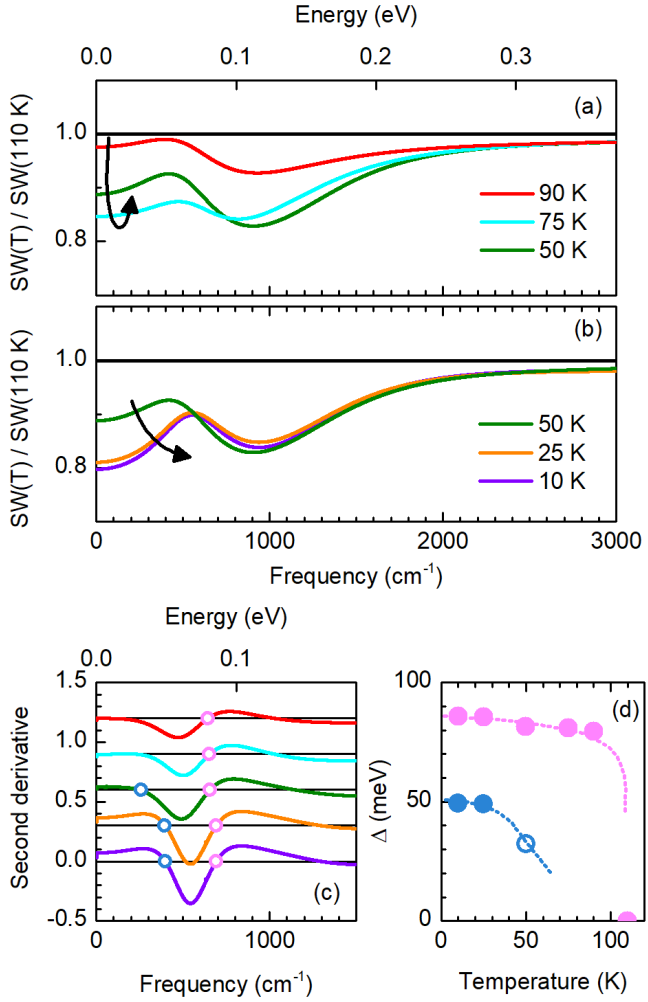}
	\caption{(a, b) Frequency dependence of SW($T$)/SW(110K) as a function of temperature. The opening of the primary gap is seen in (a), where the arrow further shows the emergence of a secondary effect at 50~K causing an increase in SW around 50~meV. The further depletion of the SW due to the emergence of a second energy scale at low energies is demonstrated with an arrow in (b). (c) Second derivative of the spectral-weight ratio (curves are shifted for clarity), with the black lines showing the zero line taken from the 90~K spectrum. The zero-crossing positions marked by circles gives a rough estimate for the energy scale of the CDW gaps. (d) Temperature dependence of both CDW gap energies. At 50~K, the second (smaller) gap is shown with an open circle, as the effect is tiny compared to the 25 and 10~K data. However, around 50~meV there is clearly an additional SW contribution at 50~K compared to 75 and 90~K.}		
	\label{energyscale}
\end{figure}
%%%%%%%%%%%%%%%%%%%%%%%%%%%%%%%%%%%%%%%%%%%%%%%%%%%%%%%%%%%%

The spectral weight (SW) is obtained as $\frac{120}{\pi}\int_0^\omega \sigma_1(\omega)d\omega$. The cutoff value of $\omega$ is chosen through the entire measured range taking into account only the interband transitions, as shown in Fig.~\ref{interband}. The SW ratio, namely, SW(T)/SW(110 K), gives the characteristics of the SW transfers (transfer direction, energy scales, etc.) in the CDW state~\footnote{Note that we choose 110 K as the reference to exclude changes in the interband absorption above \Tc.}. If there is a SW transfer from high to low energies (e.g., a narrowing of the Drude peak), the ratio is above ``1'' at low energies and then approaches unity. On the other hand, when there is a SW transfer from low to high energies, the ratio is below ``1'' until the energy transfer is completed. Here, we analyze only the spectral weight related to the interband transitions and thus directly probe the energy scales associated with the CDW.

In Fig.~\ref{energyscale}(a) and (b), the SW ratio for several temperatures below \Tc\ is plotted. The SW transfer happens in the energy range below 0.35~eV, as the ratio levels off to unity at higher energies. The SW ratios at 75 K and 90 K are qualitatively similar, while at 50 K the overall curvature increases, with an additional SW transfer observed at 10 and 25 K [panel (b)]. This transfer takes place in a much smaller energy range compared to the SW transfer caused by the primary CDW gap. It indicates an additional gap that becomes visible around 50 K. An estimate of both gaps can be given using the second derivative, as shown in Fig.~\ref{energyscale}(c). In this representation, we can also estimate approximately the main scaling below \Tc: The high energy zero crossing marks the high temperature energy scale of the gap, an additional zero crossing below 50~K is related with the additional contribution. We suggest that this second energy scale is also related to the density-wave formation and serves as the bulk probe for the multigap scenario inferred from the ARPES measurements that reported the gaps of 130~meV and 80~meV, both of them highly anisotropic across the Brillouin zone \cite{Liu2021}. Since this anisotropy is averaged out in the optical measurement, the exact gap values may differ from those in ARPES, but the qualitative behavior is remarkably similar.

%%%%%%%%%%%%%%%%%%%%%%%%%%%%%%%%%%%%%%%%%%%%%%%%%%%%%%%%%%%%%
\begin{figure*}
	\centering
	\includegraphics[width=1.65\columnwidth]{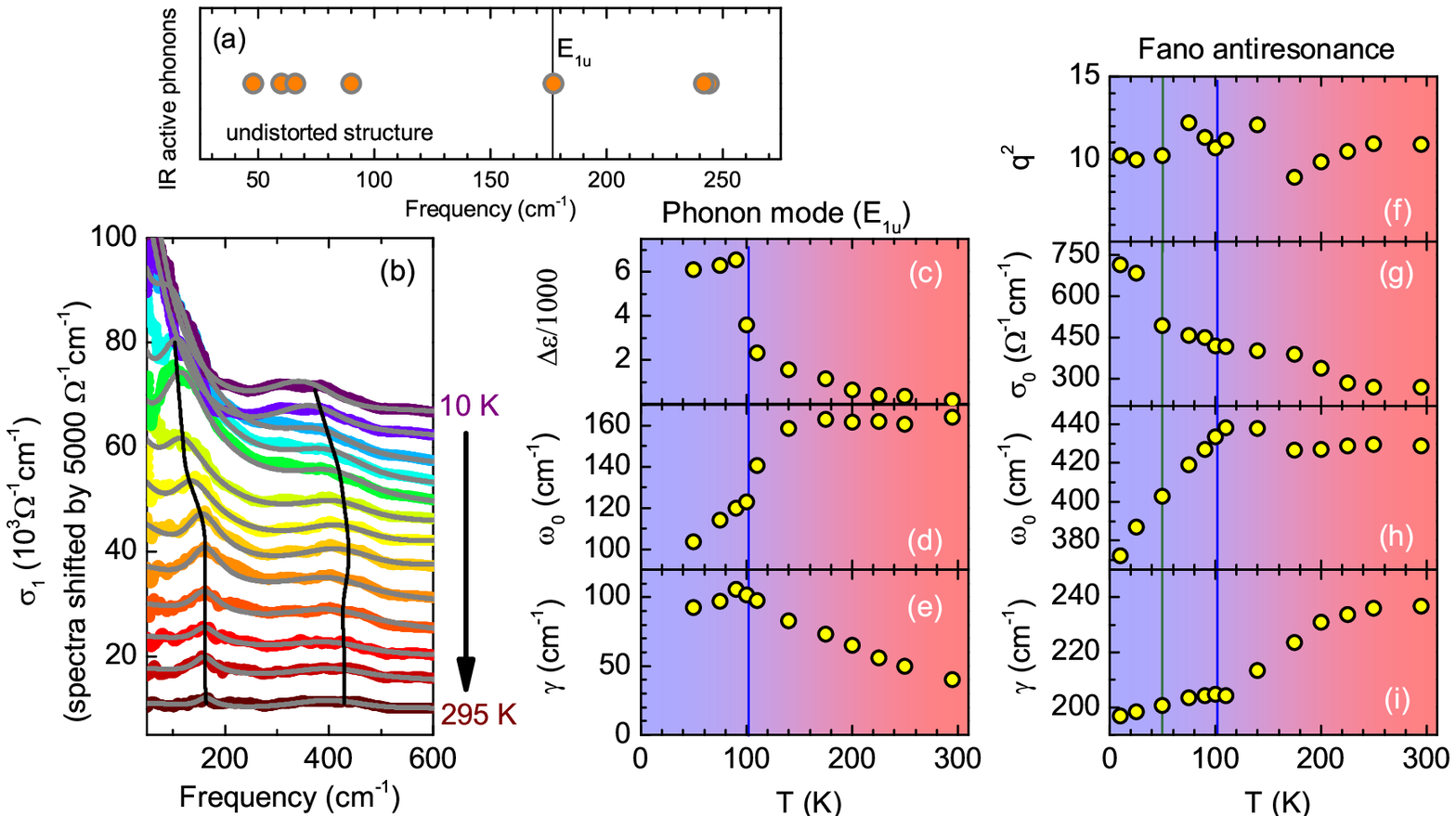}
	\caption{(a) Phonon frequencies calculated for the undistorted structure. The black line marks the E$_{1u}$ mode observed in the experimental spectra. (b) Low-energy optical conductivity, highlighting the observed phonon mode and Fano antiresonance with the solid lines. The gray lines are fits to the optical spectra. The spectra are shifted for clarity. The fit parameters obtained from the curves in (b) are given in (c-e) for the phonon mode and (f-i) for the Fano antiresonance. The blue and green lines correspond to \Tc\ and the onset temperature of the lower CDW gap, respectively.  }		
	\label{Phonon}
\end{figure*}
%%%%%%%%%%%%%%%%%%%%%%%%%%%%%%%%%%%%%%%%%%%%%%%%%%%%%%%%%%%%

Furthermore, the larger energy gap shows a rather abrupt increase right below \Tc, where the behavior is in line with the other \AVS\ systems \cite{Uykur2022, Uykur2021a}. On the other hand, the smaller energy gap seems to develop gradually with a mean-field like behavior. However, we should point out that the exact onset temperature for this second gap cannot be determined with the current measurements.

The CDW in \RVS\ is rather unusual. The opening of a density-wave gap will usually lead to an abrupt increase in the dc resistivity caused by the reduction in the density of states~\cite{Gruener1988a}. In contrast, the mere kink observed in the dc resistivity of \RVS\ at \Tc\ suggests that the mobile carriers are not significantly affected by the CDW formation. It is then plausible that the localization peak is caused by the bands in the vicinity of the $M$ point that become gapped below \Tc~\cite{Liu2021,Cho2021}, whereas the Drude peak is due to mobile carriers that reside on other bands crossing the Fermi level.

\subsection{Phonon modes}
\label{sec:Fano}

Besides these electronic features, the room-temperature spectrum shows two modes reminiscent of phonon excitations at, respectively, 160 \cm\ and 430 \cm\ [Fig.~\ref{Phonon}(b)]. While the low-energy mode is readily assigned to the IR-active E$_{1u}$ phonon according to our DFT calculations [Fig.~\ref{Phonon}(a)], the high-energy mode has the unusual Fano-like shape and cannot be interpreted in the same manner as the lower mode, because no IR-active phonons are found above 250~\cm\ in DFT [Fig.~\ref{Phonon}(a)]. Experimentally, both modes show strong anomalies across \Tc, as well as indications for a strong coupling to the electronic background.

The lower 160\,\cm\ mode can be represented with a single Lorentzian,
\begin{equation}
\sigma_1(\omega) = \frac{\Delta\epsilon\,\omega^2\omega_0^2\gamma}{4\pi[(\omega^2-\omega_0^2)^2+\gamma^2\omega^2]} \quad .
 \label{lorentz}
\end{equation}
Here, $\omega_0$, $\Delta\epsilon$, and $\gamma$ stand for the resonance frequency, intensity, and linewidth of the phonon mode, respectively. The obtained parameters in Fig.~\ref{Phonon}(c-e) reveal a strong increase in the intensity, $\Delta\varepsilon$, along with a pronounced red shift 
upon approaching the CDW transition and also below \Tc. This is contrary to the standard temperature evolution of the phonon modes that show the blue shift upon cooling as the lattice hardens.
At low temperatures, the mode is masked by the localization peak that has a strong influence on the obtained parameters. The increased broadening upon cooling is a signature of the strong electron-phonon coupling. This coupling is probably important for the appearance of the mode in the optical spectrum, because phonons in metals are usually screened by conduction electrons. Such a screening may be responsible for the absence of three other $E_{1u}$ modes, whereas three more modes (out of 7 IR-active modes in total) have the $A_{2u}$ symmetry and involve out-of-plane atomic displacements that do not couple to light in our in-plane measurement geometry.

The higher 430\,\cm\ mode is better reproduced by a Fano-like response \cite{Fano},
\begin{equation}
\sigma_1(\omega) = \frac{\sigma_0\omega\gamma[\gamma\omega (q^2-1)+2q(\omega^2-\omega_0^2)]}{4\pi[(\omega^2-\omega_0^2)^2+\gamma^2\omega^2]} \quad .
 \label{fano}
\end{equation}
The additional parameter, $q$, is the dimensionless coupling constant that describes the asymmetry of the mode and also gauges the scale of the coupling to the electronic background. We note that $q$ also takes negative values, as the high-energy mode is a strong antiresonance. The obtained parameters for the antiresonance are given in Fig.~\ref{Phonon}(f-i). A similar red-shift of the resonance frequency, $\omega_{\mathrm{0}}$, is accompanied by an increase in the intensity, $\sigma_{\mathrm{0}}$. In contrast to the low-energy mode, this feature sharpens upon cooling, which is the expected behavior with decreasing thermal effects. The coupling parameter, $q\approx-3$, remains similar across the whole temperature range and suggests that the coupling persists even below \Tc. While the anomalies at \Tc\ are clearly visible, a closer look to the variables reveals also the secondary anomalies around 50~K, such as the sudden increase in the intensity, suggesting that the antiresonance responds to changes in the electronic structure when the second CDW gap appears.

The fact that the high-energy mode cannot be assigned to a $\Gamma$-point phonon goes hand in hand with the unusual, antiresonance nature of this mode. A similar antiresonance has been observed in \KVS\ around 480~\cm, but in that case it could be regarded as an overtone of a $\Gamma$-point phonon~\cite{Uykur2022}. Such an interpretation is clearly excluded in \RVS\ where no IR-active phonon appears in the $200-220$~\cm\ range. On the other hand, such a mode could arise from a non-$\Gamma$-phonon as a result of a strong interaction with  electronic degree of freedom. Similar antiresonances have been discussed in functionalized graphene and modeled with the involvement of the $K$-point phonons \cite{Lapointe2017}. Considering the CDW formation combined with the unusual behavior of this antiresonance, phason modes could also be envisaged similar to materials with an incommensurate CDW \cite{Reagor1985, Sridhar1985, Sherwin1987, Creager1991}. However, these phason modes have been shown to appear along the out-of-plane direction and only at low energies. None of these conditions apply to our case, as the CDW order is commensurate, while the observed mode is an in-plane one and residing at a relatively high energy. Therefore, we deem the phason scenario unlikely. An interesting observation regarding the antiresonance is its prominent red shift that mirrors the red shift of the localization peak (Sec.~\ref{sec:localization}), although the former appears only below \Tc, while the latter is observed over the entire temperature range.

\subsection{Interband transitions}\label{calcsection}

%%%%%%%%%%%%%%%%%%%%%%%%%%%%%%%%%%%%%%%%%%%%%%%%%%%%%%%%%%%%%
\begin{figure}
	\centering
	\includegraphics[width=1\columnwidth]{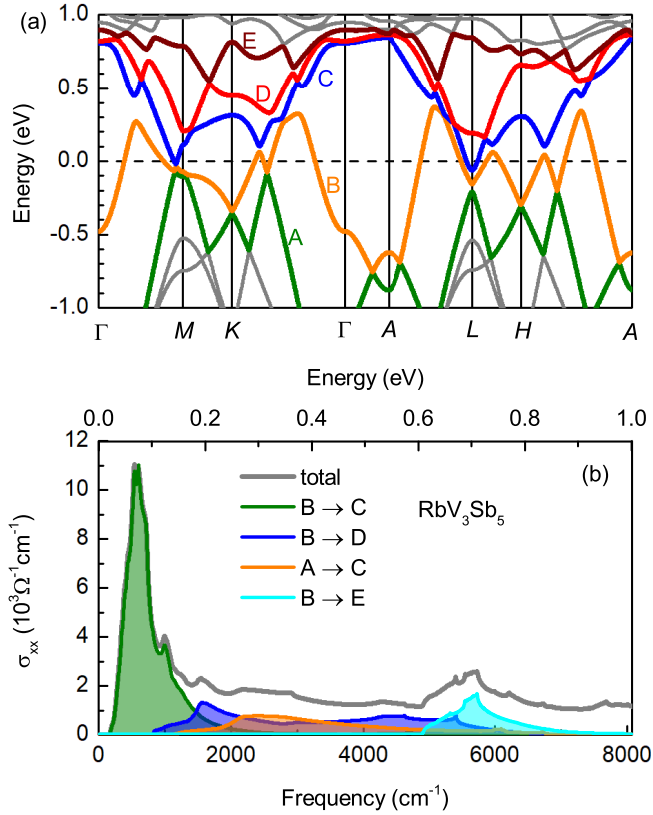}
	\caption{(a) Calculated band structure of \RVS. (b) Calculated $\sigma_{\mathrm{xx}}$ component of the optical conductivity and its band-resolved contributions. Fermi level is shifted upwards by 41~meV. }		
	\label{calculations}
\end{figure}
%%%%%%%%%%%%%%%%%%%%%%%%%%%%%%%%%%%%%%%%%%%%%%%%%%%%%%%%%%%%

%%%%%%%%%%%%%%%%%%%%%%%%%%%%%%%%%%%%%%%%%%%%%%%%%%%%%%%%%%%%%
\begin{figure}
	\centering
	\includegraphics[width=1\columnwidth]{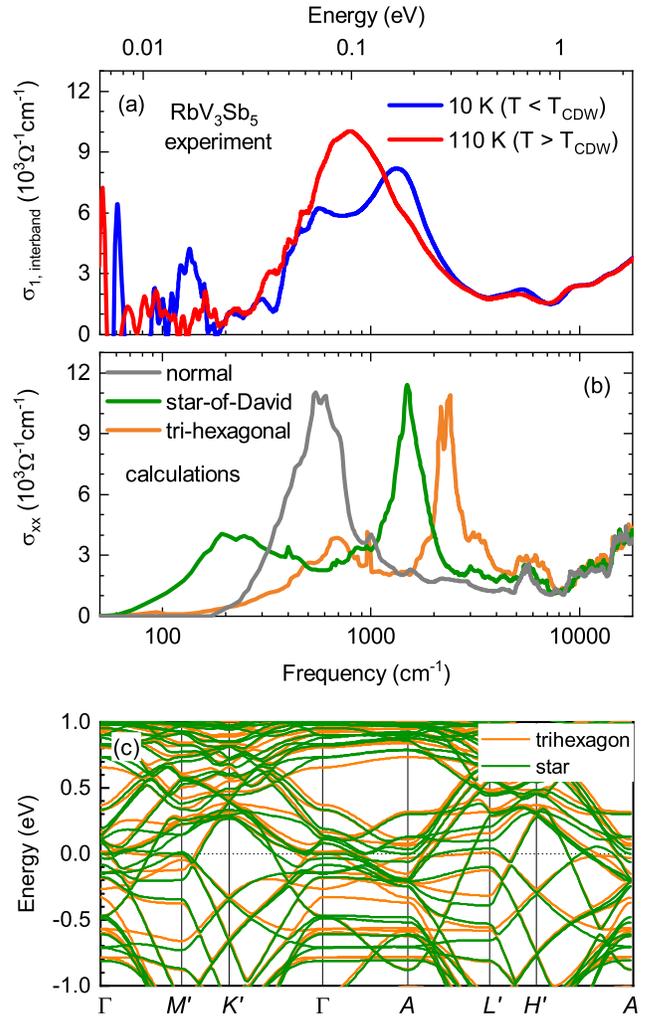}
	\caption{(a) Experimental interband transitions at 110 K (normal state) and 10 K (CDW state). (b) Calculated optical conductivity given for the normal state, as well as for the two possible types of distortions (star-of-David and tri-hexagonal). (c) Band dispersions for the CDW phases of RbV$_3$Sb$_5$ shown in the Brillouin zone of the $2\times 2$ superstructure. Note that the $M'$, $K'$, $L'$, and $H'$ points are different from the respective $M$, $K$, $L$, and $H$ points for the normal state. The $M$ of the normal state is mapped onto $\Gamma$ for the $2\times 2$ superstructure.  }		
	\label{comparison}
\end{figure}
%%%%%%%%%%%%%%%%%%%%%%%%%%%%%%%%%%%%%%%%%%%%%%%%%%%%%%%%%%%%

The interband transitions are elucidated by DFT calculations of the optical conductivity. As seen in Fig.~\ref{calculations} (b), the low-energy contributions to the in-plane optical conductivity, $\sigma_{\mathrm{xx}}$, are restricted to the transitions between bands $B$ and $C$, which occur in the vicinity of the $L$ and $M$ points of the Brillouin zone (see Fig.~\ref{calculations} (a)). In the $0.2-1.0$~eV range, the optical conductivity is dominated by three contributions that arise from the transitions $A \,\rightarrow\,C$, $B \,\rightarrow\,D$, and $B \,\rightarrow\,E$.

The experimental interband conductivity is given in Fig.~\ref{comparison} (a) below and above \Tc, after the Drude, localization, and phonon modes have been subtracted to allow for a direct comparison with the DFT results presented in Fig.~\ref{comparison} (b). We find a good agreement between experiment and calculations in the normal state when the Fermi level is shifted upwards by 41\,meV. A similar upward shift by 64\,meV was required in the case of \KVS~\cite{Uykur2022}, indicating that both compounds deviate from the band picture, and a slight renormalization of band energies is required for a proper description of their electronic structure.

For the density-wave state, two different distorted structures were considered according to the star-of-David and tri-hexagonal in-plane distortions (demonstrated in Fig.~\ref{structure}), as discussed in  \cite{Tan2021, Uykur2022, Ortiz2021a, Ratcliff2021}. The position of the additional peak in the spectrum below \Tc\ is best reproduced by the star-of-David CDW. On the other hand, this model also predicts a broad low-energy absorption feature, which we do not see in the experiment. Moreover, the peak around 70 meV in the experimental spectrum is not seen in the calculated star-of-David spectrum, however, it is reproduced by the tri-hexagonal CDW. We also note that the tri-hexagonal CDW has a higher stabilization energy of 5\,meV/f.u. (relative to the normal state) compared to 1.5\,meV/f.u. for the star-of-David CDW.

%%%%%%%%%%%%%%%%%%%%%%%%%%%%%%%%%%%%%%%%%%%%%%%%%%%%%%%%%%%%%
\begin{figure}
	\centering
	\includegraphics[width=1\columnwidth]{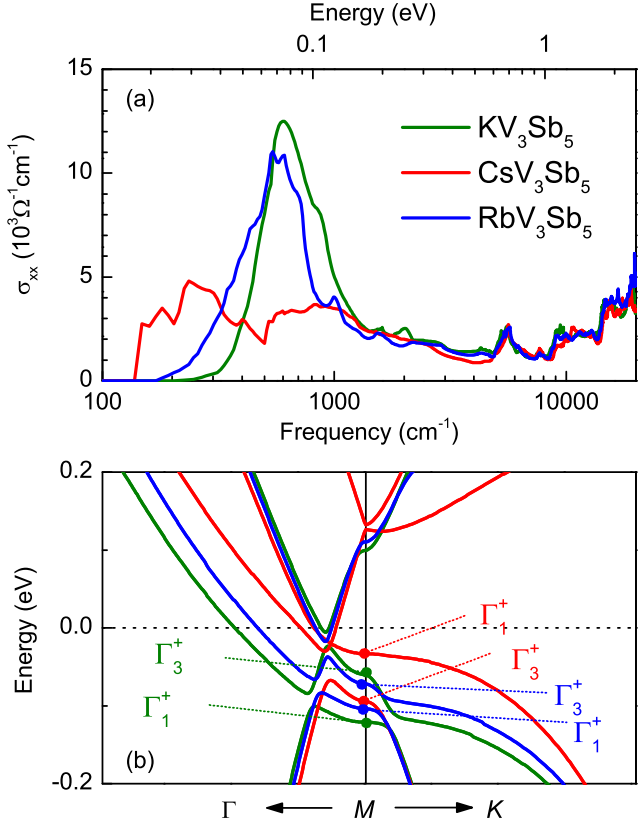}
	\caption{(a) Calculated interband optical conductivity for \KVS, \CVS, and \RVS\ in the normal state. (b) Comparison of the calculated band structures around the $M$ point. The differences in the band saddle points are clearly visible. Fermi level is shifted upwards by 41~meV and 64~meV for \RVS\ and \KVS, respectively.}		
	\label{SP}
\end{figure}
%%%%%%%%%%%%%%%%%%%%%%%%%%%%%%%%%%%%%%%%%%%%%%%%%%%%%%%%%%%%

In Fig.~\ref{comparison} (c), we show calculated band dispersions for the two possible CDW states. Both types of CDW are metallic in agreement with the persistent metallicity of RbV$_3$Sb$_5$ below \Tc. On the other hand, gap opening for some of the bands gives rise to multiple flat bands along $\Gamma-A$. The peak of the interband absorption is probably caused by optical transitions between such flat bands. Their larger separation in the tri-hexagonal case is consistent with the shift of this peak toward higher energies.

The direct comparison of the optical spectra of all three compounds of the \AVS\ series reveals important changes in the band structure and electron dynamics that appear prominently in the low-energy optical response. The interband transitions and the red-shifting localization peak are common across the whole series. Moreover, the sharp Drude peak mirrors the highly metallic nature of these systems. On the other hand, several differences need to be pointed out. (i) The Drude scattering is strongest in \RVS, causing a very broad Drude peak and a nearly flat optical conductivity in the low-energy range at high temperatures, as it overlaps with the localization peak. (ii) The low-energy interband spectrum of \RVS\ strongly resembles the one of \KVS\  with one intense absorption peak below 1000 \cm, in contrast to the three weaker peaks in \CVS\ (see Fig.~\ref{SP} (a)). This change in the interband absorption reflects the differences in the band structure related to the positions of band saddle points around $M$, as shown in Fig.~\ref{SP} (b).

With the notation used in Ref.~\cite{Park2021}, the saddle points are identified as $\Gamma_{\mathrm{1}}^{+}$ at $-105$~meV and $\Gamma_{\mathrm{3}}^{+}$ at $-70$~meV in \RVS, $-120$~meV and $-60$~meV in \KVS, and $-30$~meV and $-95$~meV in \CVS, respectively. The two saddle points are inverted in \CVS\ with respect to the \KVS\ and \RVS. Experimentally, this inversion is clearly visible as the change in the interband absorption. From electronic structure point of view, \RVS\ appears to be more similar to \KVS, yet its two saddle points become closer in energy, thus evolving toward the inverted case of \CVS.

%%%%%%%%%%%%%%%%%%%%%%%%%%%%%%%%%%%%%%%%%%%%%%%%%%%%%%%%%%%%%
\begin{figure}
	\centering
	\includegraphics[width=0.9\columnwidth]{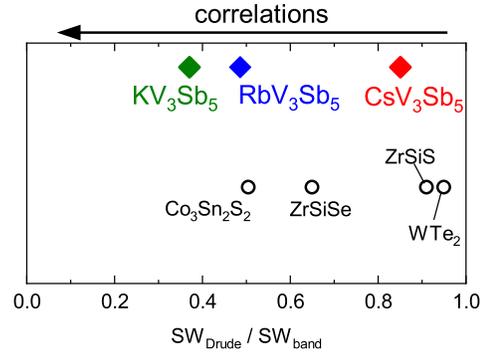}
	\caption{Ratio of the experimental and DFT-based spectral weights, SW$_\mathrm{experiment}$/SW$_\mathrm{band}$. The compounds of the \AVS\ series are shown with solid diamonds. Several topologically non-trivial Dirac/Weyl semimetals are shown for comparison (open circles) using the data from Ref.~\cite{Shao2020}.} 		
	\label{correlations}
\end{figure}
%%%%%%%%%%%%%%%%%%%%%%%%%%%%%%%%%%%%%%%%%%%%%%%%%%%%%%%%%%%%

Another interesting point is the deviation from the band picture across the \AVS\ series. It has been detected in \KVS\ and \RVS, where an upward shift of the Fermi level was required in order to reproduce interband absorption with DFT. On the other hand, no such shift was needed in the case of \CVS~\cite{Uykur2021a}. This difference can also be seen from the perspective of plasma frequencies that are compared between the experiment and DFT calculations. Previously, such a comparison of the experimental Drude spectral weight (plasma frequency squared) with the band theory was used as a gauge of electronic correlations in a series of materials, including cuprates, iron pnictides, and topologically nontrivial Dirac systems~\cite{Qazilbash2009, Shao2020}. In this comparison, the scaling parameter, SW$_\mathrm{experiment}$/SW$_\mathrm{band}$, is close to 1 for uncorrelated materials like simple metals and zero for the most correlated class of Mott insulators. Other classes of materials fall in between these two limits, with the ratios below 0.5 taken as indications for a highly correlated nature of the system.

In Fig.~\ref{correlations}, we compare the scaling factors for the \AVS\ series. Experimental plasma frequencies for \KVS\ and \CVS\ are taken from previous studies \cite{Uykur2022, Uykur2021a}. While \KVS\ is the most correlated member of the \AVS\ family, correlations seem to play only a minor role in \CVS. This result corroborates our earlier observation that a renormalization of band energies should take place in \KVS\ but not in \CVS. Finally, \RVS\ takes an intermediate position, but lies notably closer to \KVS\ and deviates from the band picture too. The similarities in the phonon anomalies and Fano antiresonance are also common to \KVS\ and \RVS. They may be associated with the moderately correlated nature of these compounds.

\section{Conclusions}

A detailed temperature-dependent optical study is presented for \RVS\ in the energy range of 10~meV--2~eV and at temperatures down to 10~K. Our data witness the highly metallic nature of the compound and the CDW transition at \Tc. The evolution of the interband absorption across the \AVS\ series mirrors changes in the band structure and especially the re-arrangement of the band saddle points (van Hove singularities) around $M$. The detailed comparison suggest a close similarity between \RVS\ and \KVS.

An unexpected feature of the low-energy optical response is the prominent localization peak that signals hindered electron dynamics. As in line with the multiband nature of the compound this localization peak coexist with the conventional Drude contribution point towards different carrier channels. In both \RVS\ and \CVS, this peak shifts to low frequencies upon cooling and merges with the Drude peak at $T\rightarrow 0$. Therefore, the localization effects in these compounds should be mostly temperature-driven and possibly related to phonons. On the other hand, the localization peak in \KVS\ remains at finite frequencies at $T\rightarrow 0$. This difference is in line with the enhanced correlation effects observed in \KVS\ (Fig.~\ref{correlations}).

Below \Tc, \RVS\ shows clear signatures of two energy scales associated with the CDW. These energy scales are in a good agreement with ARPES and confirm the bulk nature of both energy gaps that could be previously detected by surface-sensitive techniques only. We note that the agreement between the experimental and calculated optical conductivities is clearly less favorable in the CDW state than in the normal state. Each of the CDW models -- tri-hexagonal and star-of-David -- reproduces certain features of the experimental data but fails to reproduce the entire spectrum. Together with the presence of two distinct energy gaps, this may indicate the combination of multiple order parameters and the need to combine different types of distortions for a realistic modeling of the CDW in \RVS\ and in the whole \AVS\ series.

Last but not least, the prominent modes at 160~\cm\ and 430~\cm\ both reveal a strong coupling to the electronic background and put forward electron-phonon coupling as an important ingredients of the \AVS\ physics. Intriguingly, the 430~\cm\ mode with its strong antiresonance can not be interpreted as an IR-active phonon or an overtone. Moreover, such Fano-like modes appear in \KVS\ and \RVS\ that both show tangible deviations from the band picture, but not in \CVS\ where the band picture holds. The intriguing electron-phonon coupling in the \AVS\ series certainly calls for a further dedicated investigation.

Note added: recent $\mu$SR measurements on \RVS~\cite{Guguchia2022} revealed a change in the CDW state below 50~K and suggested a different type of CDW order below this temperature. It is an independent bulk probe that corroborates our conclusion on the multiple CDW gaps in \RVS. 

\begin{acknowledgments}
Authors acknowledge the fruitful discussion with Simone Fratini and the technical support by Gabriele Untereiner. We are also thankful to Berina Klis for the dc resistivity measurements. SDW and BRO gratefully acknowledge support via the UC Santa Barbara NSF Quantum Foundry funded via the Q-AMASE-i program under award DMR-1906325. BRO also acknowledges support from the California NanoSystems Institute through the Elings fellowship program. The work has been supported by the Deutsche Forschungsgemeinschaft (DFG) via DR228/51-1 and UY63/2-1. EU acknowledges the European Social Fund and the Baden-W\"urttemberg Stiftung for the financial support of this research project by the Eliteprogramme.
\end{acknowledgments}

\bibliography{RbV3Sb5}

\end{document}